

\magnification=\magstep1
\vsize=24truecm
\hsize=15.5truecm
\parskip=7pt
\baselineskip=16pt
\tolerance=5000
\font\grand=cmbx10 at 14.4truept
\def\\{{\hfil\break }}

\def\det{{\rm det}}
\def\ts{\textstyle}
\def\spa{\noalign{\smallskip}}
\def\mspa{\noalign{\medskip}}

\def\cp{{\cal P}}
\def\co{{\cal O}}
\def\la{\langle}
\def\ra{\rangle}

\pageno=0
\def\folio{
\ifnum\pageno<1 \footline{\hfil} \else\number\pageno \fi}

\rightline{STP--DIAS--92--26}

\vskip 2truecm

\centerline{\grand
Implications of an arithmetical symmetry of \break}
\centerline{\grand
the commutant for modular invariants}

\vskip 2truecm

\centerline{Ph. Ruelle}

\centerline{\it Dublin Institute for Advanced Studies}
\centerline{\it 10 Burlington Road, Dublin 4, Ireland}

\vskip 0.7truecm
\centerline{E. Thiran}

\centerline{\it Institut f\"ur Theoretische Physik}
\centerline{\it Eidgen\"ossische Technische Hochschule}
\centerline{\it H\"onggerberg, CH--8093 Z\"urich, Switzerland}

\vskip 0.7truecm
\centerline{J. Weyers}

\centerline{\it Institut de Physique Th\'eorique}
\centerline{\it Universit\'e Catholique de Louvain}
\centerline{\it Chemin du Cyclotron, 2}
\centerline{\it B--1348 Louvain--la--Neuve, Belgium}

\vskip 2truecm

\centerline{\bf Abstract}
\medskip
\leftskip 1.2truecm
\rightskip 1.2truecm
\noindent
We point out the existence of an arithmetical symmetry
for the commutant of the modular matrices $S$ and $T$.
This symmetry holds for all affine simple Lie algebras
at all levels and implies the equality of certain
coefficients in any modular invariant.
Particularizing to $\widehat{SU(3)}_k$, we classify the
modular invariant partition functions when $k+3$ is an integer
coprime with 6 and when it is a power of either 2 or 3. Our
results imply that no detailed knowledge of the commutant is
needed to undertake a classification of all modular invariants.

\leftskip=0cm
\rightskip=0cm

\vfill\eject

\noindent
{\bf 1. Introduction.}

\noindent
The classification of modular invariant partition functions
remains one of the most challenging problems in two--dimensional
conformal field theories. Various techniques have been set up to
construct modular invariants (extensions, simple currents,
automorphisms, ...), but all are lacking a completeness
criterion. A conceptual understanding of the various modular
invariants was neatly presented in [1], which puts the
aforementioned methods in a unified perspective. However, the proof
that a list of invariants is actually complete for a given
theory, is a notoriously hard question, as it involves rather
difficult linear and number--theoretic problems. As far as theories
with an affine Lie symmetry are concerned,
a complete classification is presently known only for
the affine $SU(2)$ algebra at all levels [2], and is equivalent to the
celebrated ADE classification. For
rank two algebras, a step forward was taken in [3],
where the present authors established the complete list of
invariants for theories with an affine $SU(3)$ symmetry, for all
levels $k$ such that the height $k+3$ is a prime number. In these
cases, it was proved that there are only four invariants at
each level. The proof was based on the observation that all matrices
in the commutant of $S$ and $T$ are invariant when their indices are
simultaneously multiplied by an integer coprime with $k+3$. Because of
this symmetry, the classification problem could be reduced to the
validity of a crucial arithmetical lemma. The purpose of this note
is two--fold. First, we point out that the symmetry alluded to above
is present for any algebra at any level, and we show that it forces
a series of equalities among the coefficients of any modular
invariant, physical or not. This
makes an arithmetical approach available in all cases and suggests
a possible guideline into the full classification problem.
It also provides a powerful tool to
numerically investigate high rank algebras. Following this path, we
then extend our previous result for $SU(3)$. As it happens, results
about the relevant arithmetical problem have appeared in the
mathematical litterature, which allow to classify the affine
$SU(3)$ partition functions when the height is an integer
coprime with 6, and when it is a power of either 2 or 3. We end
with some comments about higher $SU(N)$ invariants, based on
numerical results.

\vskip 0.5truecm \noindent
{\bf 2. An arithmetical symmetry of the commutant.}

\noindent
The following discussion of the commutant holds for any (untwisted)
affine simple Lie algebra, but for the sake of concreteness, we
treat in detail the unitary series.

We first recall some basic facts about the representation theory for
the chiral $\widehat{SU(N)}_k$ algebra [4]. We let the height
of the algebra be $n=k+N$. For each integer height $n \geq N$,
there are but a finite number of highest--weight unitary
representations, labelled by the strictly dominant
weights $p$ of $SU(N)$ which are in the alc\^ove
$B_n = \{p=(a_1,a_2,\ldots,a_{N-1}) \;:\; a_i \geq 1
{\rm \ and\ } a_1+\ldots+a_{N-1} \leq n-1\}$.
The number of weights in $B_n$ is equal to $n-1 \choose N-1$.
We denote the corresponding (restricted) irreducible characters
by $\chi_p(\tau)$. The weights in the alc\^ove are the \lq\lq
shifted'' weights, so that the lowest level of the affine
representation $(n;p)$ is an $SU(N)$ representation of highest
weight $\lambda=p-\rho$, where $\rho=(1,1,\ldots,1)$ is half
the sum of the positive roots.

It is well--known that the characters
transform in a unitary representation of the modular
group of the torus, $PSL(2,Z)$. Under the action of the two
generators of this group, we have $\chi_p(\tau +1)=\sum_{p' \in B_n}\,
T_{p,p'} \chi_{p'}(\tau)$ and $\chi_p({-1 \over \tau})=\sum_{p'
\in B_n}\, S_{p,p'} \chi_{p'}(\tau)$ with
$$\eqalignno{
& T_{p,p'} = e\Big({p^2 \over 2n}-{N^2-1 \over 24}\Big) \,
\delta_{p,p'}, &(1.a) \cr
& S_{p,p'} = {i^{N(N-1)/2} \over \sqrt{Nn^{N-1}}} \,\sum_{w \in W} \,
(\det\,w) \,e\Big({p \cdot w(p') \over n} \Big), &(1.b) \cr}
$$
where $W$ is the Weyl group of $SU(N)$ and $e(x)$ stands for
$\exp{(2i\pi x)}$.

The partition function of a theory with a left--right
$\widehat{SU(N)}_k$ symmetry takes the general form
$$Z(\tau,\tau^*)=\sum_{p,p' \in B_n} \, [\chi_p^*(\tau)] \, N_{p,p'}
\,[\chi_{p'}(\tau)].
\eqno(2) $$
Consistency of the theory on any torus requires the partition
function to be modular invariant [5]. The physical interpretation of
$Z(\tau,\tau^*)$ as a partition function demands in addition
that the coefficients $N_{p,p'}$ be all non--negative integers,
and that $N_{(1,1,\ldots,1),(1,1,\ldots,1)}=1$.
The classification problem is then to provide, for each value of $n$,
the complete list of all functions $Z$ (the \lq\lq physical''
modular invariants) which satisfy:
\parindent 55pt
\item{1.} $Z$ is modular invariant, \vskip -3truemm
\item{2.} the coefficients $N_{p,p'}$ are non--negative integers,
\vskip -3truemm
\item{3.} $Z$ is normalized by $N_{(1,1,\ldots,1),(1,1,\ldots,1)}=1$.
\parindent 20pt

\smallskip
The first condition requires the matrix $N$ to belong to
the commutant of $S$ and $T$:
$$ [N,S]_{p,p'} = [N,T]_{p,p'} = 0, \qquad \forall \, p,p' \in
B_n.
\eqno(3) $$
The commutation of $N$ with $T$ implies the following condition:
$$p^2 \neq p'^2 \bmod 2n \qquad \Longrightarrow \qquad N_{p,p'} = 0.
\eqno(4) $$

The commutant of $S$ and $T$ can be worked out by standard
techniques [2]. The affine characters $\chi_p$, originally
given for $p$ in $B_n$, remain well--defined on the whole weight
lattice $M^*$. Under the affine Weyl group, they transform as
$$\chi_{w(p)}=({\rm det}\,w)\,\chi_p \qquad {\rm and} \qquad
\chi_{p+nM}=\chi_p,
\eqno(5) $$
where $M$ is the co--root lattice.
Because of the second property, all weights can be taken modulo
the lattice $nM$.
The idea is then to consider the redundant set $\chi_p$ for
$p \in M^*/nM$, as if the first symmetry in (5) did not exist.
Their modular transformations are described by
simpler matrices $\hat S$ and $\hat T$ which read
$$\hat T_{p,p'}=e\Big({p^2 \over 2n}-{N^2-1 \over 24} \Big)\,
\delta_{p,p'} \quad {\rm and} \quad
\hat S_{p,p'}={i^{N(N-1)/2} \over \sqrt{Nn^{N-1}}}\,
e\Big({p \cdot p'\over n}\Big).
\eqno(6) $$
Because the alc\^ove $B_n$ is essentially $\Big(M^*/nM\Big)/W$ (up
to orbits of length smaller than $N!$ which label characters that are
identically zero), the original $S$ and $T$
matrices are recovered upon the folding with the Weyl group:
$$\eqalignno{
& S_{p,p'} = \sum_{w \in W} \,(\det \,w)\,\hat S_{p,w(p')},
\qquad {\rm for} \quad p,p' \in B_n, &(7.a) \cr
& S_{w(p),p'} = S_{p,w(p')} = (\det \,w)\,S_{p,p'}, &(7.b) \cr}
$$
and the same for $T$. In this way, the first property in (5) is
restored. Therefore, in order to compute the most
general matrix $N$ in the commutant
of $S$ and $T$, one first looks for the most general matrix $\hat N$
in the commutant of $\hat S$ and $\hat T$
and then folds it with the Weyl group, like in (7).

The construction of the commutant of $\hat S$ and $\hat T$ has
been solved in [6]. Although the commutant computed there
is not well suited for practical calculations, it readily
displays the symmetry we want to show. (For rank two algebras, a
more explicit construction of the commutant, generalizing to all
levels the simple matrices used in [3], has been given in [7].)

\medskip
We start by recalling the construction of the commutant according to
the reference [6]. Let $G_n = M^*/nM = Z_n^{N-2} \times Z_{nN}$
for $SU(N)$. ($Z_m$ denotes the congruence classes modulo $m$. Its
multiplicative group will be denoted by $Z_m^*$.) One
considers $G_n$ as a finite Hilbert space with orthonormal basis
$|p\ra$, $p \in G_n$. A set of operators $Q^p$ and $P^p$ are defined by
$$\eqalignno{
& Q^p|p' \ra = e\big({p\cdot p'\over n}\big)\,|p' \ra, \qquad
P^p|p' \ra = |p+p' \ra, &(8.a) \cr
& Q^{n\alpha} = P^{n\alpha} = 1 \quad {\rm for\ any\ } \alpha
\in M. &(8.b) \cr}
$$
The operators $\hat S$ and $\hat T$ act on $G_n$ by their matrix
representations (6). Defining now the new operators $\{k,k'\} =
e\big({k\cdot k' \over 2n}\big)\, P^k Q^{k'}$ for pairs $(k,k') \in
G_{2n} \times G_{2n}$, the key observation is that $\hat S$ and
$\hat T$ generate an $SL(2,Z)$ by their adjoint action on $\{k,k'\}$
$$ \hat S^\dagger \{k,k'\} \hat S = \{k',-k\} \quad {\rm and} \quad
\hat T^\dagger \{k,k'\} \hat T = \{k,k'-k\}.
\eqno(9) $$
$\hat S$ and $\hat T$ are represented on the pairs $(k,k')$ by
the right multiplication by $\big({0 \atop 1}{-1 \atop 0}\big)$ and
$\big({1 \atop 0}{-1 \atop 1}\big)$. Averaging the $SL(2,Z)$ action on
$\{k,k'\}$ yields operators in the commutant of $\hat S$ and $\hat T$
$$I_{\co (k,k')} = \sum_{SL(2,Z_{2nN})} \;\{ak+ck',bk+dk'\}.
\eqno(10) $$
(From the definition of $\{k,k'\}$, only the
coset $SL(2,Z_{2nN})$ acts non--trivially.) Clearly the operator
(10) depends on $(k,k')$ through its orbit $\co (k,k')$. Thus to each
orbit of $SL(2,Z_{2nN})$ on $G_{2n} \times G_{2n}$ is associated an
element of the commutant of $\hat S$ and $\hat T$. The collection of
all such elements is a generating set for the commutant.

Setting $k=l \bmod G_n$ and $k'=l' \bmod G_n$, the explicit
expression of $I_{\co (k,k')}$ depends on $(l,l')$ only, up to the
overall phase $e({k\cdot k' \over 2n})$ which we omit:
$$I_{\co (k,k')} = \sum_{SL(2,Z_{2nN})}
\, e\big({abl^2+cdl'^2+2bcl\cdot l' \over 2n}\big) \;P^{al+cl'}\,
Q^{bl+dl'}.
\eqno(11) $$
Because of (8.$b$), the sum in (11) can be partially worked out and
reduced to a sum over $SL(2,Z_{nN})$. The way this is done depends
on whether $nN$ is even or odd. When $nN$ is even, we can write
$$\pmatrix{a&b\cr c&d\cr} = \pmatrix{\alpha&\beta\cr
\gamma&\delta\cr} \, \pmatrix{1+nNs&nNt\cr nNu&1+nNs\cr}, \quad
s,t,u=0,1,
\eqno(12) $$
where the first matrix on the right--hand side of (12)
belongs to $SL(2,Z_{nN})$. The
summation over $s,t,u$ gives zero unless $Nl^2=Nl'^2=0 \bmod 2$
(automatically satisfied if $N$ is odd), in which case one obtains
$$I_{\co (k,k')}=8 \sum_{SL(2,Z_{nN})}\, e\big({\alpha\beta l^2
+ \gamma\delta l'^2 + 2\beta\gamma l\cdot l' \over 2n}\big)
\;P^{\alpha l+\gamma l'}\,Q^{\beta l+\delta l'}.
\eqno(13) $$
When $nN$ is odd, we have $SL(2,Z_{2nN}) = SL(2,Z_{nN}) \times
SL(2,Z_2)$ and the decomposition (12) changes accordingly. The
summation over the $SL(2,Z_2)$ subgroup never vanishes in this case
and yields an expression similar to (13).

{}From (8.$a$), we find that the matrix elements of $I_{\co (k,k')}$
read (numerical factors neglected)
$$\la p|I_{\co (k,k')}|p' \ra = \sum_{SL(2,Z_{nN})}
e\big({\alpha\beta l^2 + \gamma\delta l'^2 + 2\beta\gamma l\cdot l'
+ 2(\beta l+\delta l')\cdot p' \over 2n}\big) \, \delta_{p,p'+\alpha
l+\gamma l'}.
\eqno(14) $$
Let us now observe that, for any fixed integer $\nu$ coprime with
$nN$, $\big({\alpha \atop \gamma}{\beta \atop \delta}\big)$ in
$SL(2,Z_{nN})$ is equivalent to $\big({\nu^{-1}\alpha \atop
\nu^{-1}\gamma}\;{\nu \beta \atop \nu \delta}\big)$ in $SL(2,Z_{nN})$.
This change of variables allows us to replace $\alpha,\, \beta,\,
\gamma,\, \delta$ by $\nu^{-1}\alpha,\, \nu\beta,\, \nu^{-1}\gamma,\,
\nu\delta$, which in turn, is equivalent to replacing $p,p'$ by
$\nu p,\nu p'$. We thus have
$$\la p|I_{\co (k,k')}|p' \ra = \la \nu p|I_{\co (k,k')}|\nu p' \ra,
\qquad {\rm for\ any\ } \nu \in Z^*_{nN}.
\eqno(15) $$
Equation (15) is a symmetry of any matrix $\hat N$ in the commutant of
$\hat S$ and $\hat T$, since the operators $I_{\co (k,k')}$ generate it.

\medskip
The symmetry (15) has a remnant at the folded level.
Given two weights $p$ and $p'$ in $B_n$, we multiply them
by an integer $\nu$ coprime with $nN$. We get two weights $\nu p$
and $\nu p'$ which may or may not belong to $B_n$, but in any case,
after reducing them modulo $nM$ (translational part of the affine
Weyl group), there will be two unique (finite) Weyl transformations
$w_\nu$ and $w'_\nu$ which bring them
back onto two weights of $B_n$, say $p_\nu$ and $p'_\nu$. That is
$p_\nu=w_\nu(\nu p)$ and $p'_\nu=w'_\nu(\nu p')$, both in $B_n$.
(The existence of $w_\nu$ and $w'_\nu$ is guaranteed by $\nu$ being
coprime with $nN$.)
Then from the symmetry (15) and the
formulae (7) for the folding, we obtain at once that
any matrix $N_{p,p'}$ in the commutant of $S$ and $T$ satisfies
$$N_{p_\nu,p'_\nu} = (\det w_\nu)(\det w'_\nu) \, N_{p,p'}, \qquad
{\rm for\ any\ } \nu \in Z^*_{nN}.
\eqno(16) $$
This simple equation is perhaps our main result,
and has far--reaching consequences if $N_{p,p'}$ is to yield a
physical modular invariant. Indeed since $p$,
$p'$, $p_\nu$ and $p'_\nu$ are all in $B_n$, the two matrix elements
entering (16) must be positive, and hence so must be the product of
the parities of $w_\nu$ and $w'_\nu$ for {\it any} $\nu$ in
$Z^*_{nN}$. If not, the matrix elements $N_{p_\nu,p'_\nu}$ must
vanish for all $\nu \in Z^*_{nN}$.

The above arguments, including the construction of the commutant, can
be repeated verbatim for any simple Lie algebra. The main
difference lies in the structure of $M^*/nM$ as an Abelian group, which
determines which factor group of $SL(2,Z)$ is to be summed over in
(13). The height is in general defined by $n=k+h$, with $h$
the dual Coxeter number. One obtains that $M^*/nM$ is isomorphic to the
following groups
$$\eqalignno{
& A_k \;:\; Z_n^{k-1} \times Z_{n(k+1)}, &(17.a) \cr
& B_k \;:\; Z_n^{k-2} \times Z_{2n}^2 \;\hbox{ ($k$ even)  and }\;
Z_n^{k-1} \times Z_{4n} \;\hbox{ ($k$ odd)}, &(17.b) \cr
& C_k \;:\; Z_{2n}^k, &(17.c) \cr
& D_k \;:\; Z_n^{k-2} \times Z_{2n}^2 \;\hbox{ ($k$ even)  and }\;
Z_n^{k-1} \times Z_{4n} \;\hbox{ ($k$ odd)}, &(17.d) \cr
& E_6 \;:\; Z_n^5 \times Z_{3n}\;; \quad E_7 \;:\; Z_n^6 \times
Z_{2n}\;; \quad E_8 \;:\; Z_n^8, &(17.e) \cr
& F_4 \;:\; Z_n^2 \times Z_{2n}^2\;; \quad G_2 \;:\; Z_n \times
Z_{3n}. &(17.f) \cr}
$$
{}From this follows that the summation in (13) is over
$SL(2,Z_{ln})$ with the following values of $l$: $l=1$ for $E_8$;
$l=2$ for $B_{2k},\,C_k,\,D_{2k},\,E_7$ and $F_4$; $l=3$ for $E_6$
and $G_2$; $l=4$ for $B_{2k+1}$ and $D_{2k+1}$; $l=k+1$ for $A_k$.
The other minor difference is that the conditions under which
$I_{\co (k,k')}$ in (11) vanishes change, but this does not affect the
result. Therefore the relations (16) hold for any simple Lie algebra
provided we let $\nu$ vary over $Z^*_{ln}$.

\vskip 0.5truecm \noindent
{\bf 3. The parity theorem for SU(N).}

\noindent
We now make the above result more precise for the unitary series and
first indicate how the determinant factors in (16) can be computed
in those cases. Let $p=(a_1,a_2,\ldots,a_{N-1})$ be a weight of $SU(N)$.
For our purpose, the following basis is actually more convenient
than the Dynkin basis. Let $x_i=a_i+a_{i+1}+\ldots
+a_{N-1}$ for $i=1,2,\ldots,N-1$. We can write $p$ either in the
Dynkin basis (round brackets) or in the $x$--basis (square brackets):
$$p=(a_1,a_2,\ldots,a_{N-1}) \quad \longleftrightarrow \quad
\left\{\matrix{ p=[x_1,x_2,\ldots,x_{N-1}], \hfill \cr \spa
x_i=a_i+a_{i+1}+\ldots+a_{N-1}. \cr} \right.
\eqno(18) $$
The alc\^ove corresponds to
$$B_n =\{ p=[x_1,\ldots,x_{N-1}] \;:\; n > x_1 > x_2 > \ldots >
x_{N-1} > 0 \}.
\eqno(19) $$
The virtue of this basis is to make the action of the
Weyl group more transparent. If $w_i$ denotes the reflector with
respect to the $i$--th simple root, we obtain that
$$\eqalignno{
& w_i[x_1,\ldots,x_i,x_{i+1},\ldots] = [x_1,\ldots,x_{i+1},x_i,\ldots],
\quad {\rm for\ } i=1,\ldots,N-2, &(20.a) \cr
& w_{N-1}[x_1,\ldots] = [x_1-x_{N-1},x_2-x_{N-1},\ldots,
x_{N-2}-x_{N-1},-x_{N-1}]. &(20.b) \cr}
$$
Hence the first $N-2$ reflectors generate all permutations of the
$x_i$ labels. The norm of a weight $p$ is also much simpler:
$p^2 = (x_1^2 + \ldots + x_{N-1}^2) - {1 \over N}\,(x_1
+ \ldots + x_{N-1})^2$.

Let $p$ be an arbitrary weight. From the two properties,
$$\eqalignno{
& [x_1,\ldots,x_k+n,\ldots] = [x_1+n,x_2,\ldots] \bmod nM, &(21.a)
\cr
& [x_1+nN,x_2,\ldots] = [x_1,x_2,\ldots] \bmod nM, &(21.b) \cr}
$$
a set of representatives of $M^*/nM$ is obtained for $x_1 \in Z_{nN}$
and $x_k \in Z_n$ for $k=2,\ldots,N-1$. Let us write a representative
as $p=[\la x_1 \ra +jn,\la x_2 \ra,\ldots,\la x_{N-1} \ra] \bmod nM$
where $\la x \ra$ is the residue of $x$ modulo $n$, between 0 and
$n-1$, and $0 \leq j \leq N-1$. There is a unique Weyl
transformation $w$ which maps $p$ onto a weight of $B_n$ if and only
if $\la x_i \ra \neq \la x_j \ra \neq 0$ for all $i \neq j$.
We define the parity of $p$ as the determinant of $w$, $\cp(p)=\det\,w$.
The conditions on $\la x_i \ra$ ensure that $p$ is not the fixed--point
of an odd Weyl transformation, so that $\cp (p)$ is well defined.

First, if $j=0$, there is a permutation $\pi$ of the reduced
labels $\la x_i \ra$ such that $\pi (p)$ is in $B_n$. The
permutation $\pi$ is a Weyl transformation of
determinant equal to $\det \,\pi$. Second,
the Coxeter element $U=w_1\,w_2 \ldots w_{N-1}$, of determinant
$(-1)^{N-1}$, allows to bring
$j$ down to 0. Indeed, if $[\la x_1 \ra,\ldots,\la x_{N-1} \ra]$ is in
$B_n$, then
$$U(p) = [jn-\la x_{N-1} \ra,\la x_1 \ra - \la x_{N-1} \ra,
\ldots,\la x_{N-2} \ra - \la x_{N-1} \ra] \bmod nM
\eqno(22) $$
is in $B_n + [(j-1)n,0,\ldots]$. By recurrence we have $U^j(B_n +
[jn,0,\ldots]) \in B_n$.
Putting the two pieces together, we obtain for $p=[\la x_1
\ra+jn,\la x_2 \ra,\ldots,\la x_{N-1} \ra] \bmod nM$
$$\pi [\la x_1 \ra,\ldots,\la x_{N-1} \ra] \in B_n \quad
\Longrightarrow \quad U^j\pi (p) \in B_n.
\eqno(23) $$
Equation (23) easily follows from $\pi (p)=
\pi ([\la x_1 \ra,\ldots,\la
x_{N-1} \ra])+[jn,0,\ldots] \bmod nM$, a consequence of (21.$a$).
Note that (23) implies $\cp (p+[2n,0,\ldots])=\cp (p)$ for any
weight $p \in M^*$, and therefore also $\cp (p+[\ldots,0,2n,0,\ldots])
=\cp (p)$ from (21.$a$). Since $U$ is an even transformation when
$N$ is odd, we have the stronger invariance $\cp
(p+[\ldots,0,n,0,\ldots])=\cp (p)$ in this case.

We thus obtain the following algorithm to compute the parity of an
arbitrary weight
$p=[x_1,\ldots,x_{N-1}]$. First reduce the labels $x_i$ modulo
$2n$ and write $x_i = \la x_i \ra + \epsilon_i n \bmod 2n$,
with $\epsilon_i =0,1$. If
$\la x_i \ra = \la x_j \ra$ for some $i \neq j$ or if $\la x_i \ra =
0$ for some $i$, then the parity of $p$ is not defined. ($p$ would
label an affine character which is identically zero.) In all other
cases, find the permutation $\pi$ such that $n > \la x_{\pi
(1)} \ra >
\ldots > \la x_{\pi (N-1)} \ra > 0$. Then the parity of $p$ is
$$\cp (p) = (-1)^{(\epsilon_1 + \ldots + \epsilon_{N-1})(N-1)} \,
\det \,\pi.
\eqno(24) $$

With the parity of a weight given in (24), we obtain from (16)
the following criterion to decide whether a given coefficient
$N_{p,p'}$ in a physical invariant can be non--zero. From (4), we
may suppose that the norms of $p$ and $p'$ are equal modulo $2n$.
Moreover, since the
parity of $p$ depends only on its residue modulo $2n$, it is enough
to take $\nu$ in $Z^*_{2n}$ (or even in $Z^*_n$ if $N$ is odd). Note
however that, in general, $p_\nu$ really depends on the residue of
$\nu$ modulo $nN$ (remember $p_\nu \in B_n$ is the image by an
affine Weyl transformation of $\nu p$, with $p$ itself in $B_n$).

\smallskip \noindent
\hangindent=0.5cm \hangafter=1
Parity theorem for $SU(N)$.\\
{\sl Let $N_{p,p'}$ a matrix describing a physical modular invariant.\\
Suppose $p$ and $p'$ are two weights in the
alc\^ove $B_n$ such that $p^2 = p'^2 \bmod 2n$. Then for all integers
$\nu$ coprime with $nN$, we have $N_{p,p'}=N_{p_\nu,p'_\nu}$. If
the two parities $\cp(\nu p)$ and $\cp(\nu p')$ are not
equal for some $\nu$ in $Z_{2n}^*$ or $Z_n^*$ for $N$ even or odd
respectively, then $N_{p,p'}=0$.}

\smallskip
Although the above condition is weaker than the commutation of
$N$ with $S$, which must be further checked, it is nonetheless
extremely restrictive.
We have defined an action of the group $Z_{nN}^*$ on the pairs of
$B_n \times B_n$, and the theorem states that the coefficients of a
physical modular invariant are constant along each orbit.
But it is the test on the parities which makes the hard--core
of the theorem.
It says when and how the symmetry {\it may} extend, and what the
possible couplings between the characters are. Taking for example
$p'=(1,1,\ldots,1)$ labelling the character of the identity, the list
of all $p$ passing the test enumerates the primary fields
$\phi_{p}$ which may extend the affine $SU(N)$ algebra to a
larger one. As we will see in the following sections, that the
weights $\nu p$ and $\nu p'$ have the same parity for all $\nu$ in
$Z_{nN}^*$ is something rather rare, and consequently a large number
of coefficients $N_{p,p'}$ are generally required to vanish.

The relations implied by the theorem are clearly satisfied by the
diagonal invariants, but it takes a little check to show that
the parity conditions are met for the complementary invariants [8,9].
These are constructed from the outer automorphism $\mu$ of
$\widehat{SU(N)}$, defined in the Dynkin basis by
$$\mu(a_1,\ldots,a_{N-1}) = (n-{\ts
\sum}\,a_i,a_1,a_2,\ldots,a_{N-2}) = (n,0,\ldots,0) + U(p),
\eqno(25) $$
and which generates a cyclic group of order $N$, $\mu^N=1$. The
complementary invariants typically couple $p$ and $\mu^k(p)$
for some $k$. From (24), one easily checks that
$$\cp(\nu \mu^k(p)) = (-1)^{k(\nu+1)(N-1)} \, \cp(\nu p).
\eqno(26) $$
Therefore the coupling $N_{p,\mu^k(p)} \neq 0$ is compatible with
the parity test if $k(\nu+1)(N-1)$ is always even, which it is if
$N$ is odd. If $N$ is even, $\nu$ is always odd since it must be
coprime with $nN$.

Another example where the relations among the $N_{p,p'}$ can be
checked is the exceptional invariant for $SU(3)$, at height $n=24$ [10]
$$\eqalign{
E_{24}( & \tau,\tau^*) = |\chi_{(1,1)} + \chi_{(5,5)} + \chi_{(7,7)} +
\chi_{(11,11)} + \chi_{(1,22)} + \chi_{(22,1)} + \chi_{(5,14)} +
\chi_{(14,5)} \cr
& + \chi_{(7,10)} + \chi_{(10,7)} + \chi_{(2,11)} + \chi_{(11,2)} |^2
+ |\chi_{(1,11)} + \chi_{(5,7)} + \chi_{(12,1)} + \chi_{(11,12)} \cr
& + \chi_{(12,5)} + \chi_{(7,12)} + \chi_{(11,1)} + \chi_{(7,5)} +
\chi_{(1,12)} + \chi_{(12,11)} + \chi_{(5,12)} +\chi_{(12,7)}|^2. \cr}
\eqno(27) $$

Finally note that for each $\nu \in Z_{nN}^*$, the map $M_\nu \,:\,
p \rightarrow p_\nu$ is invertible on $B_n$ and so defines a
permutation of it. However the matrix $(M_\nu)_{p,p'}$ does not
qualify to describe a physical invariant unless $\nu=-1$. Indeed the
map $M_{-1}$ is just the charge conjugation $C$ acting by
$C(a_1,a_2,\ldots,a_{N-1})=(a_{N-1},a_{N-2},\ldots,a_1)$.

\vskip 0.5truecm \noindent
{\bf 4. The case of SU(3) when n is coprime with 6.}

\noindent
The equation (24) gives the parity of any weight in the $x$--basis.
For $SU(3)$ however, the expression is just as easy in the Dynkin
basis. If $p=(\la a \ra,\la b \ra) \bmod n$, then
$$\cp(p)=\cases{
+1 & if $\la a \ra+\la b \ra < n$, \cr \mspa
-1 & if $\la a \ra+\la b \ra > n$. \cr}
\eqno(28) $$
If we use the affine Dynkin basis, writing $\tau=(a,b,n-a-b)$, then
the equation (28) is equivalent to $\cp(\tau)=+1$ or $-1$ according
to whether $\la a \ra + \la b \ra + \la n-a-b \ra = n$ or $2n$.
This makes it clear that the parity is invariant under any
permutation of the affine Dynkin labels. For this reason, it is
better to use the affine weights, that we generically denote by
$\tau$. $\tau$ is in $B_n$ if its three labels are integers between
1 and $n-1$.

The parity theorem of Section 3 is precisely what was used in [3] to
classify the modular invariants of $SU(3)$ when $n$ is a prime
number. Choosing $\tau=(1,1,n-2)$, it was proved that
for no weight $\tau'$ in $B_n$ are the parities $\cp(\nu \tau)$ and
$\cp(\nu \tau')$ equal for all $\nu$, except for the trivial
solutions, namely $\tau'$
is a permutation of $\tau$. (The equality of the norms of the two
weights was not imposed.) Although in a totally different context, the
parity theorem for $SU(3)$, in its full power, was in fact
investigated by Koblitz and Rohrlich some fifteen years ago [11].
Their main result is as follows.

\noindent
{\sl Theorem [11].} Let $n$ an integer coprime with 6. Let
$\tau=(a,b,n-a-b)$ and $\tau'=(a',b',n-a'-b')$ two weights of $B_n$.
Then the parities of $\nu \tau$ and $\nu \tau'$ are equal for every
$\nu \in Z^*_n$ if and only if $\tau'$ is a permutation of $\tau$.

\smallskip
This remarkable result allows the classification
for the corresponding heights without looking any further
into the details of the commutant.

{}From the above
theorem, the character of the identity representation $\chi_{(1,1)}$
can only couple to itself, $\chi_{(1,n-2)}$ and $\chi_{(n-2,1)}$.
The last two possibilities are readily excluded because the norms of
the corresponding weights are not equal modulo $2n$ to that of
$(1,1)$. Therefore, any
partition function looks like $Z(\tau,\tau ^*)=|\chi_{(1,1)}|^2 +
\ldots$, which shows that the affine $SU(3)$ symmetry does not
extend. As to the other weights, the above theorem and the norm
condition imply that a weight $p$ can only couple to one of the
following possibilities: $p$ itself,
$C(p)$, $\sigma(p)$ or to $C\sigma(p)$, where
$\sigma(p)=\mu^{nt(p)}(p)$, $C$ is the charge conjugation, $\mu$ is
the outer automorphism (25) and $t(p)$ is the triality of $p$. This
is a local result, valid for each weight $p$ taken separately, but
it is not difficult to show that these couplings must be global as
well (see reference [12] for more details). We thus obtain the final
result that there are only four modular invariant partition functions,
the diagonal and the complementary found in [8], given by
$$ N_{p,p'} = \delta_{p',p}, \quad {\rm and} \quad
N_{p,p'} = \delta_{p',\sigma(p)},
\eqno(29) $$
and their $C$--twisted version, obtained by replacing $N_{p,p'}$
by $N_{p,Cp'}$. For $n=5$, the second invariant of (29) is identical to
the $C$--conjugate of the first one.

\vskip 0.5truecm \noindent
{\bf 5. Powers of 2 and 3 for SU(3).}

\noindent
Koblitz and Rohrlich also examined the parity theorem when $n$ is a
power of 2 or 3. We start with the powers of 2, which is the simplest
case. As in the theorem we used in Section 4, they do not impose
any condition on the norms of $\tau$, $\tau'$.
When $n$ is a power of 2, we do impose such conditions
in order to make the statements
simpler. We say that the norms of $\tau$ and $\tau'$ do not match
modulo $2n$ if $p^2 \neq p'^2 \bmod 2n$ for any choice of 2--label
({\it i.e.} non--affine)
weights $p$ and $p'$ obtained respectively from $\tau$ and $\tau'$.

\medskip
Set $n=2^m \geq 16$. The following result has been proved in [11].
Let $\tau=(a,b,n-a-b)$ and $\tau'=(a',b',n-a'-b')$ two weights of
$B_n$ with matching norms. Suppose in addition that
${\rm gcd}(\tau,\tau')=1$. Then the parities of $\nu \tau$ and
$\nu \tau'$ are equal for every $\nu \in Z^*_{2^m}$ if and only
if $\tau$ and $\tau'$ are either permutations of each other, or
else permutations, modulo $n$, of $u(1,1,n-2)$ and $u({n \over 2}-1,
{n \over 2}-1,2)$ for some $u \in Z^*_{2^m}$.

We proceed as follows. First this theorem shows that $(1,1)$
can only couple to itself and to $({n \over 2}-1,{n \over 2}-1)$, which
signals a possible extension of the symmetry by the field
$\phi_{({n \over 2}-1,{n \over 2}-1)}$. If $(1,1)$ does not couple
to $({n \over 2}-1,{n \over 2}-1)$, {\it i.e.}
$N_{(1,1),(n/2-1,n/2-1)}=0$ hence $N_{u(1,1),u(n/2-1,n/2-1)}=0$ by
the parity theorem, then the $\widehat{SU(3)}$ symmetry
does not extend and, using the same argument as in Section 4,
the only invariants are the diagonal and the complementary (same as
in (29)), plus
their $C$--conjugates. On the other hand, if $(1,1)$ does couple
to $({n \over 2}-1,{n \over 2}-1)$, then the invariant is
necessarily of the form $Z(\tau,\tau^*) =
|\chi_{(1,1)}+\chi_{(n/2-1,n/2-1)}|^2 + \ldots$ (because of (16) with
$\nu={n \over 2}-1$), and thus involves an extension of the symmetry.
We show that this is not compatible with modular invariance unless $n=8$.

In order to do this, we look at another part of $N$. From the above
theorem, the weight $(\epsilon,n-3)$, with $\epsilon=n \bmod 3$,
only couples to itself and its conjugate. The corresponding
$2 \times 2$ blocks of $N$ and $S$ must commute, which yields
$$N_{p,p'} = \pmatrix{\alpha & \beta \cr \beta & \alpha \cr},
\qquad {\rm for} \quad p,p' \in \{(\epsilon,n-3),\,(n-3,\epsilon)\},
\eqno(30) $$
which is also a consequence of (16). If we now enforce the commutation
$[N,S]_{p,p'}=0$ for $p$ in $\{(1,1),({n \over 2}-1,{n \over
2}-1)\}$ and $p'$ in $\{(\epsilon,n-3),(n-3,\epsilon)\}$, we find
that there is no solution for $\alpha,\beta$ unless the equation
$\sin{2\pi \over n} - \sin{6\pi \over n}=0$ holds, which it does
for $n=8$ only. Hence for $n \geq 16$, there is no extension of the
symmetry and the only physical invariants are the diagonal
and the complementary. For $n=8$, it is easy enough to check by
hand that there is one exceptional invariant, given by
$$\eqalign{
E_8(\tau,\tau^*) = |\chi_{(1,1)} & +\chi_{(3,3)}|^2 + |\chi_{(1,3)}
+\chi_{(4,3)}|^2 + |\chi_{(3,1)}+\chi_{(3,4)}|^2 \cr
& + |\chi_{(1,4)}+\chi_{(4,1)}|^2 + |\chi_{(2,3)}+\chi_{(6,1)}|^2
+ |\chi_{(3,2)}+\chi_{(1,6)}|^2. \cr}
\eqno(31) $$
The invariant (31) is the reduction of the diagonal invariant
of $\widehat{SU(6)}$ level 1, into which $\widehat{SU(3)}$ level 5 is
conformally embedded [10]. Thus for all heights $n=2^m \geq 16$,
there are four modular invariants in terms of the unrestricted
characters, and they are given in (29). There are six invariants for
$n=8$, and only two for $n=4$.

\bigskip
Finally, we consider the powers of 3, $n=3^m \geq 9$. This case is
slightly more difficult since the symmetry always extends.
We start by giving the relevant result from [11].

Let $\tau=(a,b,n-a-b)$ and $\tau'=(a',b',n-a'-b')$ two weights of $B_n$.
Then the parities of $\nu \tau$ and $\nu \tau'$ are equal for every $\nu
\in Z^*_{3^m}$ if and only if $\tau$ and $\tau'$
are either permutations of each other, or else they are permutations,
modulo $n$, of $u(3^k,{n \over 3}-2 \cdot 3^k,{2n \over 3}+3^k)$
and $u(3^{k+1},{n \over 3}-2 \cdot 3^k,{2n \over 3}-3^k)$ for some
$k$ between 0 and $m-2$ and some $u \in Z^*_{3^{m-k}}$.

{}From this result and the equality of the norms, we obtain that the
three weights $(1,1),\,(n-2,1),\,(1,n-2)$ can only be coupled among
themselves, and the same is true for the two weights (1,2) and
(2,1). Imposing $[N,S]_{p,p'}=0$ for $p,p'$ running over these five
weights leads to the following two situations.
Either the symmetry does not extend and the only invariants are the
diagonal and its $C$--twisted version [12], or else the symmetry
extends and the partition function looks
like $Z(\tau,\tau^*)=|\chi_{(1,1)}+\chi_{(n-2,1)}+\chi_{(1,n-2)}|^2 +
\ldots$. In the second case, $Z(\tau,\tau^*)$ must be expressible in
terms of the combinations $\tilde\chi_{(a,b)} =
\chi_{(a,b)}+\chi_{(n-a-b,a)}+\chi_{(b,n-a-b)}$ and $\chi_{({n
\over 3},{n \over 3})}$. Moreover the three weights labelling the
characters in $\tilde\chi_{(a,b)}$ must have the same norm modulo
$2n$, implying that $\tilde\chi_{(a,b)}$ cannot appear if $(a,b)$
is a weight with a non--zero triality.

The remaining characters, $\tilde\chi_{(a,b)}$ for $(a,b)$ a root and
$\chi_{({n \over 3},{n \over 3})}$, are the reduced characters of an
extended theory $\cal T$, possessing a larger symmetry than
$\widehat{SU(3)}$. ($\cal T$ contains three different
characters $\tilde\chi_{({n \over 3},{n \over 3})}^i$, $i=1,2,3$, which
all reduce to the same affine character.) According to the results of
[1], every partition function of $\cal T$ originates from an automorphism
$\sigma$ of the fusion rules of $\cal T$ and is necessarily of the form
(the sum also includes the $\tilde\chi_{({n \over 3},{n \over 3})}^i$)
$$Z(\tau,\tau^*) = \sum_{(a,b) \in B_n \cap M} \;
[\tilde\chi_{(a,b)}^*(\tau)] \,\, [\tilde\chi_{\sigma(a,b)}(\tau)].
\eqno(32) $$
Furthermore $\sigma$ is a permutation that has to commute with the
matrix $\tilde S$ of the extended theory.

Let us show that the extra couplings, namely those for which the
affine weights $\tau$ and $\tau'$ are not permutations of each other,
must be excluded. Suppose the contrary, namely that
the partition function is $Z(\tau,\tau^*)=\ldots \; +
\tilde\chi_{u(3^k,n/3-2\cdot 3^k)}^* \,\,
\tilde\chi_{u(3^{k+1},n/3-2\cdot 3^k)} + \ldots$ for some $k$ and
some $u$. (The only other possibility
is the same coupling with one of the two weights conjugated,
but clearly these two cases are both compatible with modular
invariance or none of them is. Note that instead of $u(3^k,{n \over
3}-2\cdot 3^k)$ and $u(3^{k+1},{n \over 3}-2\cdot 3^k)$, we should
take their residues modulo $n$, themselves in $B_n$. It makes no
difference in the following.) From the parity theorem, the above
coupling cannot depend on $u$, so we may take
$u=1$. Because $Z$ must be of the form
(32), the automorphism $\sigma$ which exchanges
$p_1=(3^k,{n \over 3}-2\cdot 3^k)$ and $p_2=(3^{k+1},{n \over 3}-2
\cdot 3^k)$ must leave the extended $\tilde S$ matrix invariant. In
particular, one must have $\tilde S_{(1,1),p_1}=\tilde S_{(1,1),p_2}$.
But $(1,1),\,p_1$ and $p_2$ are roots (different from $({n \over
3},{n \over 3})$) and thus these matrix elements
of $\tilde S$ are just three times the same matrix elements of $S$.
So we obtain the same condition on the original $S$ matrix,
$S_{(1,1),p_1}=S_{(1,1),p_2}$, which explicitly reads
$$\sin{2\pi ({n \over 3}-3^k) \over n} - \sin{2\pi 3^k \over n}=
\sin{2\pi ({n \over 3}+3^k) \over n} - \sin{2\pi 3^{k+1} \over n}.
\eqno(33) $$
By using the identity $\sin{({2\pi \over 3}+x)} + \sin{x} =
\sin{({2\pi \over 3}-x)}$, the equation
(33) simplifies to $\sin{2\pi 3^{k+1} \over n}=0$, which is
impossible on account of the inequalities $0 \leq k \leq m-2$.
Therefore all these \lq exceptional' couplings are ruled out.

At this stage, we have that each extended character
$\tilde \chi_{(a,b)}$ can couple to itself or to $\tilde \chi_{(b,a)}$.
This means that, in (32), the automorphism
$\sigma$ maps $(a,b)$ onto itself or onto $(b,a)$, for each root
$(a,b)$ separately. However, if it is to commute with $S$, the
charge conjugation $C$ can only act globally, even when it is, like
here, restricted to act on the roots of $B_n$ only. (To obtain this,
one shows that, for $n=3^m$, the matrix element $S_{(1,4),(a,b)}$ is
real if and only if $a=b$, $b=n-2a$ or $a=n-2b$.) Thus
the automorphism $\sigma$ is either the
identity or $C$, leading to two and only two invariants with
an extension of the symmetry (they were already found in [9]):
$$Z(\tau,\tau^*) = \sum_{\scriptstyle (a,b) \in B_n \cap M
\atop \scriptstyle (a,b) \neq ({n \over 3},{n \over 3})} \;
|\tilde\chi_{(a,b)}(\tau)|^2
\;+\; 3\,|\chi_{({n  \over 3},{n \over 3})}|^2
\eqno(34) $$
and its $C$--conjugate. Putting everything together, we obtain, for
$n=3^m \geq 9$, four modular invariant partition
functions in terms of the unrestricted affine
characters. Clearly, for $n=3$, there is only one invariant.

\vskip 0.5truecm \noindent
{\bf 6. Numerical investigation for higher ranks.}

\noindent
In this last section, we report on some numerical results concerning
higher $SU(N)$ algebras. The parity theorem of Section 3 is
particularly well suited for numerical
studies as the computations are at all times performed on a few integer
variables ($n$, $\nu$ and the Dynkin labels). It requires no large
memory, unlike a systematic search of all physical invariants. We
emphasize that the list of couplings $N_{p,p'}$ allowed by the
parity theorem does not classify the modular invariants. In particular,
the commutation with $S$ must be separately checked. Despite this,
its real power is to reveal the heights where something special may
be expected.

It is straightforward to run a computer program that examines the
parity theorem. It only needs the formulae for the norm and
the parity of a weight, both easier in the $x$--basis (see Section 3).
To test the parities,
we note that we can let $\nu$ range from 1 to $n$ or to $n \over 2$
for $N$ even or odd respectively, on account of the following
property of the parity: $\cp((2n-\nu)p)=(-1)^{N(N-1)/2}\, \cp(\nu
p)$ for $N$ even, $0 < \nu < n$, and $\cp((n-\nu)p) =
(-1)^{N(N-1)/2}\,\cp(\nu p)$ for $N$ odd, $0 < \nu < {n \over 2}$.

By choosing $p'=\rho=(1,1,\ldots,1)$ in the parity theorem, we have
looked for heights $n$ at which an extension of the symmetry can be
expected. For $SU(3)$ ($n \leq 500$), $SU(4)$ ($n \leq 150$) and
$SU(5)$ ($n \leq 100$) we have listed the weights $p$ in the
alc\^ove such that $p^2=\rho^2={N(N^2-1) \over 12} \bmod 2n$ and
such that the parities of the $\nu$--multiples of $\rho$ and $p$
coincide for all $\nu$. The results are more conveniently
expressed in terms of the orbits of the weights under the
automorphism $\mu$ of equation (25) and the charge conjugation $C$.
Such orbits are in general of length $2N$. We will say that two
orbits are allowed by the parity theorem to couple if each orbit
contains one weight allowed to couple with at least one
weight in the other orbit. So we look for the orbits which can
couple to the orbit of the identity $\rho$.

\smallskip
In addition to its self--coupling,
the orbit of the identity for $SU(3)$
can also couple to that of $({n \over 2}-1,{n \over 2}-1)$ when $n$ is
divisible by 4. It is easy to check from (24) or (28) that the
parity test is satisfied, while the equality of the norms requires
$n=0 \bmod 4$. The surprising outcome of our numerical
computation is
that these \lq regular' couplings are the only ones allowed by the
parity theorem for $n$ up to 500, except in two cases, $n=24$ and $n=60$.

For $n=24$, the identity can couple to (the orbits of) (1,1), (5,5),
(7,7) and (11,11). All these
couplings remain allowed by the commutation with $S$, and lead
to the exceptional invariant of equation (27). It was numerically
checked in [13] that there is no other exceptional invariant at that
height. The case $n=60$ is in some respects similar to $n=24$. We find
that the identity can couple to (1,1), (11,11), (19,19) and to
(29,29). However here none except the self--coupling
survives the commutation with $S$, so that there
is no exceptional extension of the symmetry.

In all other cases, namely $n \neq 24,\,60$,
the only new extension of the symmetry can only come from
$({n \over 2}-1,{n \over 2}-1)$ when $n$ is divisible by 4.
Using the numerical fact that $(1,4)$ can
only couple to its own orbit (not true for $n=24$), we have checked
that the extension by $({n \over 2}-1,{n \over 2}-1)$ is not
compatible with the commutation with $S$ unless $n=8$ or $n=12$,
where there is indeed an exceptional invariant $E_8$ (given in
(31)) and $E_{12}$ [10].

{}From this, we conclude that for $n \leq 500$, the only new modular
invariants can only be of the kind found in [1] at $n=12$, {\it i.e.}
associated to automorphisms of the theory extended by the primary
fields $\phi_{(1,n-2)}$ and $\phi_{(n-2,1)}$ at height $n=0
\bmod 3$. Combined to the results of [12], our numerical
calculations prove that the list of
invariants of [10] is complete for $n$ not divisible by 3 and
smaller or equal to 500.

Before leaving $SU(3)$, we would like to make the following remark.
It may be desirable to study the parity theorem without imposing the
norm requirement, as was done in [11]. However, numerical evidence
shows that the resulting classification of couplings (even to (1,1)
only) is bound be
something rather complicated when $n$ is even. For example, at $n=42$,
no less than 12 orbits can couple to the identity if the norm matching
condition is dropped, whereas only one remains (the identity orbit
with itself) if that condition is reinstalled. Strangely, the
norm condition seems to play almost no r\^ole when $n$ is odd. This
pattern strengthens for higher ranks.

Our next example is $SU(4)$ with $n \leq 150$. Here too we find one
series of regular couplings when $n$ is even: the orbit of the
identity can couple to that of $({n \over 2}-2,1,{n \over 2}-2)$
provided $n \neq 4 \bmod 8$ (for norm reasons). These couplings
appear in the exceptional invariants coming from
conformal embeddings, at $n=8,\,10$ [14] and $n=12$ [15]. Apart
from this regular series, the situation much depends on whether $n$
is even or odd, as indicated above. When $n$ is odd, we find just one
case where the identity can couple to another orbit: at $n=15$, with
the orbit of (1,3,4). If $n$ is even, there are
additional couplings to the identity for most even heights up to 90.
(They are not many though: their maximal number is 4, attained at
$n=30$.) In the range from 92 to 150 and presumably onwards,
the only allowed couplings are
those of the above regular series. We have not investigated the
question as to whether they yield acceptable extensions.

Finally in the case of $SU(5)$,
there are three series of regular couplings, which again
appear when $n$ is even: $(1,{n \over 2}-2,{n \over 2}-2,1)$ (no
condition on $n$ from the norm), $({n \over 2}-3,1,1,{n \over 2}-3)$
(for $n=0 \bmod 4$) and $({n \over 2}-3,2,2,{n \over 2}-3)$
(also for $n=0 \bmod 4$). (It is straightforward although tedious to
check that they all pass the parity test for any $n$.)
As for $SU(3)$ and $SU(4)$, the couplings
belonging to these regular series are involved in the exceptional
invariants coming from conformal embeddings, in this case at $n=8$,
$n=10$ [14] and $n=12$ [15].
As to the other couplings, the situation is very much like in
$SU(4)$. When $n$ is odd, there are just two heights with additional
couplings: at $n=15$, the identity can couple to (1,3,4,4), and
at $n=17$, it can couple to (3,3,5,4). For $n$ even, there are
generally (many) more couplings (16 more couplings for $n=42$).

\smallskip
In conclusion, we see that the combined requirements from the norm
condition and especially the parity test put severe constraints
on the way the characters can be coupled to form a physical modular
invariant. This is even more true when the height is an odd integer.
The numerical results show that for the $SU(N)$
series, there seems to be a considerable difference between the even
and odd heights, as already well illustrated by $SU(2)$ and $SU(3)$.
We note that all conformal embeddings of $\widehat{SU(N)}_k$
occur at even heights [16], and that the conjectured list of
exceptional invariants due to automorphisms of an extended algebra
also contains even heights only [17].
We are not aware of the existence of an exceptional invariant at
an odd height, and our results indicate that nothing exceptional is
to be expected there.

\vskip 0.5truecm \noindent
{\bf 7. Conclusion.}

\noindent
We have shown that, for any affine simple Lie algebra at any level,
the (unfolded) commutant of the modular matrices $\hat S$ and
$\hat T$ possesses an
arithmetical symmetry. Any matrix in the commutant is invariant when
its two indices are simultaneously multiplied by an integer coprime
with $ln$, $n$ being the height (level plus dual Coxeter number) and
$l$ is an algebra dependent integer. At the folded level, this
symmetry results in a
parity theorem, which says that certain coefficients of a modular
invariant must be equal, and gives a condition under which these
coefficients must vanish.

This result has two main virtues. It first shows that a
precise account of the commutant is not really needed. The parity
theorem precisely embodies what we believe is its most important
property. Instead, the
classification of modular invariants is reduced to the study of the
parity theorem, which is a purely arithmetical problem. Secondly,
it opens the possibility of a much wider numerical investigation,
allowing to study high rank algebras at large levels, a thing which
was so far beyond computational feasability.

Using available results relevant to the parity theorem,
we classified the $SU(3)$ partitions functions when the height is
coprime with 6, and when it is a power of 3 or 2.

\vskip 1.5truecm \noindent
{\bf References.}

\item{[1]} G. Moore and N. Seiberg, {\it Nucl. Phys.} B313
(1989) 16.
\item{[2]} A. Capelli, C. Itzykson and J.B. Zuber, {\it Commun.
Math. Phys.} 113 (1987) 1.
\item{[3]} Ph. Ruelle, E. Thiran and J. Weyers, {\it Commun. Math.
Phys.} 133 (1990) 305.
\item{[4]} V. Kac, {\it Infinite Dimensional Lie Algebras},
Birkh\"auser, Boston 1983.
\item{[5]} J. Cardy, {\it Nucl. Phys.} B270 (1986) 186.
\item{[6]} M. Bauer and C. Itzykson, {\it Commun. Math. Phys.}
127 (1990) 617.
\item{[7]} Ph. Ruelle, Ph.D. thesis, September 1990.
\item{[8]} D. Altsch\"uler, J. Lacki and P. Zaugg, {\it Phys. Lett.}
205B (1988) 281.
\item{[9]} D. Bernard, {\it Nucl. Phys.} B288 (1987) 628.
\item{[10]} P. Christe and F. Ravanini, {\it Int. J. Mod. Phys.}
A4 (1989) 897.
\item{[11]} N. Koblitz and D. Rohrlich, {\it Can. J. Math.} XXX
(1978) 1183.
\item{[12]} Ph. Ruelle, {\it Automorphisms of the affine $SU(3)$
fusion rules}, to appear.
\item{[13]} D. Verstegen, {\it Nucl. Phys.} B346 (1990) 349.
\item{[14]} A.N. Schellekens and S. Yankielowicz, {\it Nucl. Phys.}
B327 (1989) 673, {\it Nucl. Phys.} B334 (1990) 67.
\item{[15]} G. Aldazabal, I. Allekotte, A. Font and C. N\'u\~nez,
{\it Intern. J. Mod. Phys.} A7 (1992) 6273.
\item{[16]} A.N. Schellekens and N.P. Warner, {\it Phys. Rev.} D34
(1986) 3092;\\
F.A. Bais and P.G. Bouwknegt, {\it Nucl. Phys.} B279 (1987) 561.
\item{[17]} D. Verstegen, {\it Commun. Math. Phys.} 137 (1991) 567.

\end